\documentclass[english]{article}
\usepackage[T1]{fontenc}
\usepackage[latin9]{inputenc}
\usepackage{amssymb}
\usepackage{graphicx}
\usepackage{esint}
\usepackage{subscript}

\makeatletter
\newcommand{\lyxaddress}[1]{
	\par {\raggedright #1
	\vspace{1.4em}
	\noindent\par}
}

\makeatother

\usepackage{babel}
\begin{document}
\title{Dynamical correlations in simple disorder and complex disorder liquids}
\author{Bernarda Lovrin\v{c}evi\'{c}$^{\ddagger}$\thanks{bernarda@pmfst.hr},
Martina Požar$^{\ddagger},$ Ivo Juki\'{c}\thanks{Doctoral School of Biophysics, University of Split, Croatia}$^{\dagger}{}^{,\ddagger}$
\\
David Perera$^{\nmid}$ and Aurélien Perera$^{\dagger}$\thanks{corresponding author (aup@lptmc.jussieu.fr)}}
\maketitle

\lyxaddress{$^{\dagger}$Laboratoire de Physique Théorique de la Matière Condensée
(UMR CNRS 7600), Sorbonne Université, 4 Place Jussieu, F75252, Paris
cedex 05, France.}

\lyxaddress{$^{\ddagger}$Department of Physics, Faculty of Science, University
of Split, Ru\dj era Boškovi\'{c}a 33, 21000, Split, Croatia.}

\lyxaddress{$^{\nmid}$ Laboratoire de Traitement et Communication de l'Information,
Telecom Paris, Institut Polytechnique de Paris, 19, place Marguerite
Perey CS 20031 F-91123 Palaiseau, France.}
\begin{abstract}
Liquids in equilibrium exhibit two types of disorder, simple and complex.
Typical simple disorder liquid are liquid nitrogen, or weakly polar
liquids. Complex liquids concern those who can form long lived local
assemblies, and cover a large range from water to soft matter and
biological liquids. The existence of such structures leaves characteric
features upon the atom-atom correlation functions, concerning both
atoms which directly participate to these structure and those who
do not. The question we ask here is: does these features have also
characteristic dynamical aspects, which could be tracked through dynamical
correlation functions. Herein, we compare the van Hove function, intermediate
scattering function and the dynamical structure factor, for both types
of liquids, using force field models and computer simulations. The
calculations reveal the paradoxical fact that neighbouring atom correlations
for simple disorder liquids relax slower than that for complex disorder
liquids, while prepeak features typical of complex disorder liquids
relax even slower. This is an indication of the existence of fast
kinetic self-assembly processes in complex disorder liquids, while
the lifetime of such assemblies itself is quite slow. This is further
confirmed by the existence of a very low-$k$ dynamical pre-peak uncovered
in the case of water and ethanol.
\end{abstract}

\section{Introduction}

While liquids are known to be the archetype of disorder (excluding
all liquid crystalline materials), liquid nitrogen and water belong
to two different classes of liquids \cite{Perera2016,Pozar2016}.
The first one is characterised by simple disorder, while the second
is about complex disorder, due to the presence of hydrogen connectivity
\cite{Dixit2002,Hbond_Guo2003}, which gives a local coherence that
simple disorder liquids do not have. Other complex disorder liquids,
such as alcohols, show chain like clustering of the hydroxyl head
groups \cite{EXP_Warren1933,EXP_Pierce1938,EXP_MD_Benmore2000,EXP_Matija_alcohols,EXP_SIM_Matija_Butanol,Perera2007}.
These are elementary forms of self-assembly, which is the major property
of more complex liquids, such as soft matter \cite{EXP_Microemulsions1,EXP_Microemulsions2,EXP_Teubner_Strey1987}
or biological liquids \cite{Kundu2018}. Admittedly, disorder is not
really a physical chemistry property, such as energy or entropy. However,
it is related to the structural properties, which are a window into
the microscopic arrangement of particles, from statistical point of
view. In that, it is much more specific of the nature of liquids than
properties like energy or entropy. 

Perhaps the typical experimental evidence of such complex form of
disorder is the radiation scattering pre-peak \cite{EXP_Warren1933,EXP_Pierce1938,EXP_Magini1982,EXP_Narten_MethEth,EXP_Finns_alcohols,EXP_Sarkar_Methanol,EXP_Sarkar_ethanol,EXP_Matija_alcohols,EXP_Alcohols_Karmakar1999}
in the scattering intensity $I(k)$, which witnesses the existence
of local aggregates. For typical liquids of small molecules, the relevant
experiment is wide angle x-ray scattering, which covers the range
from $k=0$ until the atom-atom peak which about $k\approx1.5-2\mathring{A}^{-1}$.
Interestingly, not all complex disorder systems have a radiation scattering
pre-peak \cite{EXP_EthWat_Nishikawa1993,EXP_PropWAT_Akiyama2004,EXP_PropWat_Takamuku2004,Almasy2019}.
This is because the scattering pre-peak is the result of sum of pre-peaks
and anti-peaks in specific atom-atom structure factors \cite{Almasy2019,Pozar2020},
which sometimes tend to exactly cancel each other, thus leaving a
net small-$k$ raise in $I(k)$, which is interpreted as the presence
of large heterogeneity \cite{Anisimov}.

$I(k)$ is however a static indicator, and it would be interesting
to have information about the kinetics and lifetime of the local heterogeneity
and labile structures, and also how they affect various dynamical
quantities. Unfortunately, there is no such thing as time dependent
radiation scattering intensity $I(k,t)$ for x-ray scattering. Spectroscopy
methods provide informations about the molecular relaxation, but these
mostly focus on intra-molecular relaxation, and the coupling with
inter-molecular relaxation must be obtained through interpretations.

Computer simulation techniques can provide a direct access to atom-atom
dynamical pair correlation functions and the corresponding structure
factors. However, this implies a necessary model dependence, which
can be a drawback in some cases, such as that of mixtures involving
aggregate forming components, as exemplified with the difficulties
in computing the Kirkwood-Buff integrals \cite{Perera2004,Ploetz2011,Gupta2012,Ganguly2013,Krueger2013,Milzetti2018,Dawass2018}.
Despite these difficulties, it is interesting to obtain dynamical
correlation functions, since these could provide a good idea of the
molecular kinetics, as well an opportunity to check agreements with
dynamical quantities, such as transport coefficients for example.

In the present work, we compute atom-atom van Hove functions $G(r,t)$,
corresponding intermediate scattering functions $F(k,t)$ and dynamical
structure factors $S(k,\omega)$, hence covering the direct and reciprocal
spaces, as well time and frequency domains, and for various types
of single component liquids, both of simple and complex disorder types.
The liquids studied herein are carbon-tetrachloride, acetone, water
and ethanol. In addition to allowing us to find correspondances between
typical structural features associated with local clustering in various
representation of correlations, we expect to learn correspondence
between the relaxation mechanisms themselves, since these functions
are microscopic probes of better investigation power than $I(k,t)$
could access.

There has been previous report of dynamical correlations in the literature,
which we briefly review here. It is interesting to note that not,
compared to calculation from computer of static correlations covering
many types of liquids, there are remarkably much less coverage with
dynamical ones. Early calculations concern mostly simple liquids,
and lately water, but in the supercooled regime. Going into details,
dynamical structure factors have been calculated for liquid metals,
such as rubidium \cite{Rahman1974,Yoshida1978} and aluminum \cite{Sjogren1979,Ebbsio1980},
and molten salts \cite{Kahol1977}. Similar analysis have been performed
for polyisoprene and their MD simulations gave a clearer interpretation
of the experimentally observed results \cite{Moe1999}. Models for
glass forming systems, like silica, were studied via MD simulations
by Sciortino \emph{et al.} \cite{SciortinoHandle2019} and for liquid
iron by Wu \emph{et al.} \cite{Wu2018}. Dynamical correlations have
lately been reported from inelastic x-ray scattering for liquid water
at ambient conditions and inelastic neutron scattering at various
temperatures \cite{Noguere2021}. MD results of the coherent dynamic
structure factor of liquid water at the mesoscale have been presented
by Alvarez \emph{et al.} \cite{Alvarez2021}.

In view of this scarcity of coverage, the present study could, although
focused on the quality of disorder, could be equally seen as providing
the missing coverage other types of liquids than those studied so
far.

The remainder of the manuscript is as follows. We first recall what
dynamical atom-atom correlations are, and how they are interrelated.
We then explain how they are calculated and the models and simulation
details. Then, in the results section, we study into some detail these
correlation functions and investigate what they can tell us about
the molecular relaxations at different levels. A final section gathers
what perspectives this study inspires and outlines our conclusions.

\section{Atom-atom dynamical correlations}

We make the choice of dealing with atom-atom correlation functions,
even in the case of molecular liquids, instead of orientational correlation
functions. It is generally believed that these latter functions are
a better choice, since they contain information irretrievably lost
in the former \cite{Kezic2011}. In addition, atom-atom correlation
functions can be obtained from the orientational ones, but not the
opposite \cite{Gray_Gubbins}. However, these orientation correlations
cannot be manipulated directly and necessitate and infinite number
of terms in the standard rotational invariant expansion \cite{Blum2003}.
For instance, the atom-atom functions can be obtained only as a sum
of such infinite number of terms \cite{Textbook_Hansen_McDonald}.
This raises practical issues about the convergence, which can become
crucial when specific orientations play an important, such as the
hydrogen bond, for instance. Since atom-atom correlation functions
contain both the intra-molecular and inter-molecular correlations,
we believe that these are much simpler functions to calculate, and
they always come in a finite number, even when large proteins are
concerned. There are additional two appealing reasons for using atom-atom
functions. Firstly, they implicitly assume that molecular liquids
are like a ``soup of atoms'', some of which are grouped through
intra-molecular correlations, hence making a link between covalent
binding and labile binding, with open interesting possibilities to
unify both representations into a single one. Secondly, as shown below
in this Section, the intra-molecular part is in fact very naturally
connected to the self part of the van Hove function, thus becoming
a necessary ingredient in a dynamical approach of molecular liquids,of
which the static representation is only the $t=0$ limit. We believe
that these 2 reasons enforce an overwhelming bias in choosing atom-atom
correlations over orientational ones.

For the above mentioned reasons, we dwell into details of these dynamical
functions from the atomic perspective, even though many of these details
can be found in most text books \cite{Textbook_Hansen_McDonald,RISM_Hirata2003}.

\subsection{The van Hove function}

The time dependent microscopic density per particle of atomic species
$a$ , at position $\mathbf{r}$ and time $t$, is defined as:
\begin{equation}
\rho_{i;a}(\mathbf{r},t)=\delta\left[\mathbf{r}-\mathbf{r}_{i;a}(t)\right]\label{rho_ia}
\end{equation}
where $\mathbf{r}_{i;a}(t)$ is the time dependent position of particle
$i$ or atomic species $a$, in the lab fixed frame. This definition
holds regardless whether the atom belongs to a molecule or not, since
we use the ``molecular liquid=soup of atoms'' convention, where
each atomic species is named differently, even if it concerns the
same atom type. In this convention, the two hydrogen atoms of water
are labeled differently (e.g $H_{1}$ and $H_{2}$, for instance).
From this per-particle microscopic density, one defines the microscopic
random variable which is the total density of atomic species $a$
by summing over all atom $i$ of species $a$:
\begin{equation}
\rho_{a}(\mathbf{r},t)=\sum_{i}\rho_{i;a}(\mathbf{r},t)\label{rho_a}
\end{equation}
From these two random variables, one can compute various type of statistical
averages in selected statistical ensembles. For instance, in the constant
$NVT$ Canonical ensemble, the simple 1-body average give the trivial
relation:
\begin{equation}
<\rho_{a}(\mathbf{r},t)>=\frac{N_{a}}{V}=\rho_{a}\label{rho_a_av}
\end{equation}
where $N_{a}$ is the number of atoms of species a in the volume V,
$\rho_{a}$the number density of species a, and we have used the fact
that we consider liquids at equilibrium which are homogeneous and
isotropic. In the ``molecular liquid=soup of atoms'' convention,
for given molecular species $m$ , since all atoms are uniquely represented,
the number density $\rho_{a_{m}}$ of any atomic species $a_{m}$
within that molecule is exactly that $\rho_{m}$ of the molecular
species itself $\rho_{a}=\rho_{m}$. For this reason, we will use
$\rho_{a}$without any ambiguity about its meaning.

Two body correlations can be defined in a similar way, such as $\rho_{ab}^{(2)}(\mathbf{r}_{1},\mathbf{r}_{2},t)=<\rho_{a}(\mathbf{r}_{1},t)\rho_{b}(\mathbf{r}_{2},0)>$
which correlates two types of atoms $a$ and $b$ at respective positions
$\mathbf{r}_{1}$ and $\mathbf{r}_{2},$and taken at two different
times, where we set arbitrarily the origin of time the system of atomic
species $b$. Using the fact that one can use the variable change
$(\mathbf{r},\mathbf{r}')\rightarrow(\mathbf{r},\mathbf{R})$, where
$\mathbf{r}=\mathbf{r}_{1}-\mathbf{r}_{2}$ and $\mathbf{R}=(\mathbf{r}_{1}+\mathbf{r}_{2})/2,$whose
Jacobian is $1$, using the random variable in Eq.(\ref{rho_ia})
we define the \emph{dimensionless} self van Hove function for homogeneous
and isotropic liquids as:
\begin{equation}
G_{aa}^{(s)}(r,t)=\frac{1}{\rho_{a}^{2}V}\sum_{i}\int d\hat{\mathbf{r}}\int d\mathbf{R}<\rho_{i;a}(\mathbf{r}_{1},t)\rho_{i;a}(\mathbf{r}_{2},0)>\label{Gs_aa}
\end{equation}
where we have integrated both on the irrelevant $\mathbf{R}$ variable,
as well as the unit vector $\hat{\mathbf{r}}$ representing the orientational
part of $\hat{\mathbf{r}}$ of $\mathbf{r}$, since the system is
isotropic and the pair correlation function depends only on $r=|\mathbf{r}|=|\mathbf{r}_{1}-\mathbf{r}_{2}|$
. 

Similarly, using the total microscopic density defined in Eq.(\ref{rho_a}),
we define the \emph{dimensionless} total van Hove function as:

\begin{equation}
G_{ab}(r,t)=\frac{1}{\rho_{a}\rho_{b}V}\int d\hat{\mathbf{r}}\int d\mathbf{R}<\rho_{a}(\mathbf{r}_{1},t)\rho_{b}(\mathbf{r}_{2},0)>\label{G}
\end{equation}
Using Eq.(\ref{rho_a}), this latter equation can be split into two
parts, a distinct van Hove correlation function
\begin{equation}
G_{ab}^{(d)}(r,t)=\frac{1}{\rho_{a}\rho_{b}V}\int d\hat{\mathbf{r}}\sum_{i\neq j}\int d\mathbf{R}<\rho_{i;a}(\mathbf{r}_{1},t)\rho_{j;b}(\mathbf{r}_{2},0)>\label{Gd}
\end{equation}
and a \emph{new} self van Hove function:
\begin{equation}
G_{ab}^{(s)}(r,t)=\frac{1}{\rho_{a}\rho_{b}V}\sum_{i}\int d\hat{\mathbf{r}}\int d\mathbf{R}<\rho_{i;a}(\mathbf{r}_{1},t)\rho_{i;b}(\mathbf{r}_{2},0)>\label{Gs_ab}
\end{equation}
which differs from that in Eq.(\ref{Gs_aa}) since it can now concern
two atoms which do not belong the same species. However, since, by
definition, the self correlation can only concern similar atoms, we
see that one can now consider as self part, the correlations between
different atoms, but belonging to the \emph{same} molecular species.
In this case, the self van Hove function is nothing else that the
intra-molecular dynamical correlation function. Needless to say, $G_{ab}^{(s)}(r,t)=0$
when atoms $a$ and $b$ belong to different molecules. Eq.(\ref{Gs_aa})
is contained in Eq.(\ref{Gs_ab}), and we will from now on consider
Eq.(\ref{Gs_ab}) as the definition of the self van Hove function.

Several remarks can be made. First of all, these 3 functions hold
for neat liquids as well for mixtures, which is a convenient unified
description of both cases. Secondly, since we have defined dimensionless
van Hove functions, the two following static limits hold:
\begin{equation}
G_{ab}^{(d)}(r,t=0)=g_{ab}(r)\label{g=00003DGd(t=00003D0)}
\end{equation}
where $g_{ab}(r)$ is the usual static pair distribution function
between atoms $a$ and $b$, and 
\begin{equation}
G_{ab}^{(s)}(r,t=0)=w_{ab}(r)\label{w=00003DGs}
\end{equation}
where $w_{ab}(r)$ is the intra-molecular correlation function which
appears in the RISM theory.

Thirdly, two types of dynamical Kirkwood-Buff integrals (dKBI) and
running dKBI (RdKBI) can be defined, one for the distinct function
and one for the self function. We first define the RdKBI as
\begin{equation}
K_{ab}(r,t)=4\pi\int_{0}^{r}dss^{2}\left[G_{ab}^{(d)}(s,t)-1\right]\label{dKBI}
\end{equation}
\begin{equation}
K_{ab}^{(s)}(r,t)=4\pi\int_{0}^{r}dss^{2}G_{ab}^{(s)}(s,t)\label{RdKBI}
\end{equation}
The dynamical KBI are then defined as $dKBI(t)=K_{ab}(\infty,t)$
and the self part as $dKBI^{(s)}(t)=K_{ab}(\infty,t)$. It turns out
that both these functions are time independent in ergodic equilibrium
liquids. Since the self van Hove function represent the correlation
of one atom with itself across time, in the infinite time limit, in
an ergodic equilibrium liquid, the trajectory of any atom should cover
the entire system. Hence its integral is just the volume occupied
by the atomic species, which is the molar volume $V_{a}$ scaled by
the mole fraction $x_{a}$ for the molecular species containing atom
$a$:
\begin{equation}
K_{ab}^{(s)}(\infty,t)=x_{a}V_{a}\delta_{ab}\label{KBI-self}
\end{equation}
Similarly, for ergodicity reasons, the distinct dynamical correlation
function, $K_{ab}(\infty,t)$ is the same as the static KBI:
\begin{equation}
K_{ab}(\infty,t)=K_{ab}\label{KBI}
\end{equation}
It is important to underline that both dynamical KBI are the same
for all pairs of atoms belonging to same molecular species pairs.

\subsection{The intermediate scattering function}

The intermediate scattering functions are the spatial Fourier transforms
of the van Hove functions, and come in same 3 varieties, total, self
and distinct, and are dimensionless functions, generically defined
as:
\begin{equation}
F^{(\bullet)}(k,t)=\sqrt{\rho_{a}\rho_{b}}\int d\mathbf{r}\exp(i\mathbf{\mathbf{k.r}})G^{(\bullet)}(k,t)\label{Fkt}
\end{equation}
where the superscript $(\bullet)$ could be either blank for the total
function, or $(d)$ for the distinct function and$(s)$ for the self
function.

Interesting equalities occur for the static case. The self-part reduces
to the Fourier transform of the RISM theory w-matrix elements
\begin{equation}
F_{ab}^{(s)}(k,t=0)=w_{ab}(k)=j_{0}(kd_{ab})\label{Fs-wk}
\end{equation}
where the second equality holds in case of rigid molecules, with $d_{ab}=|\mathbf{r}_{a}-\mathbf{r}_{b}|$
is the distance between the two atoms $a$ and $b$ inside the molecule,
and with $j_{0}(x)=\sin(x)/x$ being the zeroth-order spherical Bessel
function.

The total part reduces to the total structure factor:
\begin{equation}
F_{ab}(k,t=0)=w_{ab}(k)+\sqrt{\rho_{a}\rho_{b}}\int d\mathbf{r}\exp(i\mathbf{k.r})\left[g_{ab}(r)-1\right]\label{Fkt-Debye}
\end{equation}
the latter which appears in the Debye expression for scattering intensity
\cite{Debye1915,Debye1954}. We note that, for like atoms, this expression
reduce to that of the usual static structure factor
\begin{equation}
S_{aa}(k)=1+\rho_{a}\int d\mathbf{r}\exp(i\mathbf{k.r})\left[g_{aa}(r)-1\right]\label{Sk}
\end{equation}
The relations to the dynamical KBI are directly derived from above
relations and Eqs.(\ref{KBI-self},\ref{KBI}):

\begin{equation}
F_{ab}^{(s)}(k=0,t)=V_{a}\label{Fsk=00003D0}
\end{equation}
\begin{equation}
F_{ab}^{(d)}(k=0,t)=K_{ab}\label{Fk=00003D0}
\end{equation}

\subsection{The dynamical structure factor}

The dynamical structure factors are the time-Fourier transforms of
the 3 types of van Hove functions, generically defined as:
\begin{equation}
S_{ab}^{(\bullet)}(k,\omega)=\int_{-\infty}^{+\infty}dt\exp(i\omega t)F_{ab}^{(\bullet)}(k,t)\label{Skw}
\end{equation}
where the ($\bullet$) stands for any of the (t,d,s) symbols. 

The dynamical structure factors are interesting from several point
of view. Firstly and most importantly, they appear naturally in the
theoretical approaches, such as the Mori-Zwanzig formalism, since
the operational form involved both the $r$ Fourier transform and
the time Laplace transform. The resulting function $S(k,z)$ is related
to the dynamical structure factor $S(k,\omega)$ through the well
known relation \cite{Textbook_Hansen_McDonald}:
\begin{equation}
S^{(\bullet)}(k,\omega)=\lim_{\epsilon\rightarrow0}\frac{1}{\pi}\mbox{Re}\left[S^{(\bullet)}(k,z=\omega+i\epsilon)\right]\label{z2w}
\end{equation}
We will not expand further on these aspects in this work in the present
context.

Secondly, dynamical structure factors allow to make contact with molecular
hydrodynamics. This is perhaps the most intriguing aspect of liquids,
that the small $k$-range from k$=0$ to the main atom-atom peak around
$k\approx1.5-2\mathring{A}^{-1}$ covers in fact the entire spatial
range from atom size to macroscopic size. In computer simulation with
system size $L$, the largest k value corresponds to $k_{L}=2\pi/L$.
For typical size $L\approx40\mathring{A},$this gives $k_{L}\approx0.12$,
which is close enough to k=0. In other words, it is not unexpected
that hydrodynamic modes, such as the Rayleigh and Brillouin sound
modes could be covered within such tiny simulated scales. With simulation
times of $t=100$ps, the smallest frequency is $\omega\approx0.0$6Ghz
, which is much smaller than $\omega\approx0.8-1.2$Ghz where the
Brillouin peak is found. In other words, one could extract sound speed
for model molecular liquids by finding if the corresponding low $k$
low $\omega$ peak is obtained from calculated $S(k,\omega)$.

Finally, it is $S(k,\omega)$ which is directly obtained from scattering
experiments, mostly through light scattering. Such experiments, however,
cannot detect small molecules such as water. They are more appropriate
for nano-sized molecules, hence present little interest in this work.

\subsection{Dynamical scattering intensities}

It is straightforward to extend the Debye formula for the static scattering
intensity $I(k)$ \cite{Debye1915,Debye1954} to dynamical equivalents
$I(k,t)$ and $I(k,\omega)$. This way, one can report calculated
intensities for x-ray and neutron experiments, using the static form
atomic factors $f_{a}$, even if no real experiment can obtain these
intensities at present. For x-ray scattering in a single component
system, one has:

\begin{equation}
I(k,t)=r_{0}^{2}\rho\sum_{a,b}f_{a}(k)f_{b}(k)F_{ab}(k,t)\label{Ikt}
\end{equation}
\begin{equation}
I(k,\omega)=r_{0}^{2}\rho\sum_{a,b}f_{a}(k)f_{b}(k)S_{ab}(k,\omega)\label{Ikt-1}
\end{equation}
where $r_{0}=0.2818\mathring{A}$ is the electronic radius and $\rho$
the molecular number density.

\section{Molecular models, simulation details and methodology}

The simulations were performed with the GROMACS program package \cite{MD_Gromacs}.
For each system we followed the same protocol. The initial random
configurations of 2048 molecules were obtained by the Packmol program
\cite{MD_Packmol}, which were subsequently energy minimized and equilibrated
in the NPT ensemble for 1 ns. NPT production runs of nearly 1 ns were
used to collect 10 000 independent configurations for the calculation
of all statistical properties. The liquids were simulated in ambient
conditions of T = 300 K and p = 1 bar, maintained by the Nose-Hoover
thermostat \cite{MD_Nose,MD_Hoover} and Parinello-Rahman barostat
\cite{MD_Parrinello_Rahman_1,MD_Parrinello_Rahman_2}. The former
algorithm had the time constant of 0.1 ps, whereas the latter one
had the time constant of 1 ps. The integration algorithm of choice
was leap-frog REF, with the time step being 1 fs. The cut-off radius
for short-range interactions was 1.5 nm. For the long-range Coulomb
interactions we employed the particle mesh Ewald (PME) method \cite{MD_PME},
with FFT grid spacing of 0.12 nm and interpolation order of 4. The
constraints were handled with the LINCS algorithm \cite{MD_LINCS}.
The forcefields used were: SPC/E for water \cite{FF_SPCe}, OPLS-UA
for ethanol \cite{FF_OPLS_alcohols}, OPLS-AA for carbon tetrachloride
\cite{FF_OPLS_CCl4} and a modified OPLS-UA for acetone. The latter
forcefield is based on the original OPLS-UA model by Jorgensen and
coworkers \cite{FF_OPLS_UA_Acetone}, which we modified to better
reproduce some thermodynamic and dynamic quantities. The details of
the forcefield modification are documented in detail in previous publications
\cite{Pozar2020_acetone,Pozar2020_nonpolar_dynamics}.

\subsection{Numerical evaluation of the dynamical functions}

The LiquidLib package \cite{MD_LiquidLib} was used to extract the
total $G_{ab}(r,t)$ and self $G_{ab}^{(s)}(r,t)$ atom-atom van Hove
functions, directly from the Gromacs trajectory file. We have modified
the code in order to obtain dimensionless functions, and to include
the unlike atom self van Hove function Eq.(\ref{Gs_ab}) among the
usual like atom self van Hove function Eq.(\ref{Gs_aa}). This is
an important step, since it allows to separate the unlike atom distinct
van Hove function, which, in turn, allows to fully understand separately
the intra-molecular and inter-molecular dynamics of all atoms in molecular
liquids.

The LiquidLib package allows to sample the van Hove functions in the
same $r$-grid as the static correlation function $g_{ab}(r)$, but
with the user's choice for the time grid. In order to avoid data storage
burden, we have computed the time dependence in 3 different samplings.
The first sampling consists in 10 time-points from 0 to 1ps, spaced
by 0.1ps. The second grid is from 0 to 10ps, spaced by 1ps, and the
final grid from 0 to 100ps, spaced by 10ps. This is motivated by the
3 facts. Firstly, the time decays are quite fast and all functions
are significantly decayed to their limits ($1$ for the total function
and $0$ for the self part) by 100ps. However, this is not true for
small r values, specially those within the core part. The reason for
this is detailed in the Results section below. Secondly, most interesting
changes occur within the first ps, and the second grid compensates
for any slower dynamics occurring beyond 1ps. Lastly, the sampling
near $r=0$ are quite noisy and demand excessively long runs. This
is more serious at small times than longer ones. In order to minimise
computational times, we have fitted the initial time decays to a Gaussian
function. 

While the $r$-Fourier transforms are made exactly as for the static
case, using fast Fourier techniques as in our previous work \cite{Pozar2016,Pozar2020},
the time-Fourier transforms require interpolating the time dependence
from the different 3 time grids discussed above. We have used a time
grid of 0.1ps. In addition, care must be taken for time functions
for $r$-values inside the core, since these are not decayed by t=100ps.
In such case, we have used the following trick. We use the observed
empirical property that all time correlations decay exponentially
at large times, specifically those in the 3rd window between 10ps
and 100ps. The functions are fitted to an exponential decay which
covers mostly the large times. Then the difference between the data
and the exponential fit is a short ranged function which can be numerically
Fourier transformed. Then, the Lorentzian, corresponding to the exact
Fourier transform of the exponential, is added to the numerical transform,
in order to obtain the total time Fourier transform.

For large $r$ values, the van Hove data is generally noisy for all
times. Interestingly, in all such cases, the global $t$-decay is
found to be exponential. We therefore replaced these functions by
their exponential fit. When the fitting does not cover properly the
small times, which happens in the intermediate $r$-values, then only
the tail region is replaced by an exponential. This way the entire
procedure can be automated.

\section{Results}

We are interested in reporting details of various dynamical correlation
functions, and find details which can relate to their local structural
and relaxation features.

\subsection{Simple disorder liquids}

In this section we present dynamical correlations for carbon tetrachloride
and acetone. Both are polar liquids, but CCl\textsubscript{4} is
considered less polar than acetone. Both liquids have polar order,
but they do not form labile clusters such as that we consider for
complex liquids. For each liquid we present the self part and the
distinct or total part of the dynamical correlation functions, for
typical 2 pairs of sites. For each function, we represent all the
29 times, covering the range {[}0,1ps{]}, {[}0,10ps{]} and {[}0,100ps{]},
with 10 points in each interval. We observe that the times decays
are always monotonous, with no-crossover. Hence, the curves in the
figures are not labeled.

\subsubsection{Apolar or weakly polar liquid: Carbon tetrachloride}

CCl\textsubscript{4} is a polar molecule, but the force field model
uses small partial charges, which mimic the fact that the molecule
is polarisable. The central carbon has a charge of +0.248, while the
Cl sites share equally the opposite charge. In terms of charge ordering,
it become noticeable only when the charge is above 0.6 \cite{Perera2015}.
Therefore CCl\textsubscript{4} can be considered as simple disorder
liquid.

\begin{figure}
\includegraphics[scale=0.45]{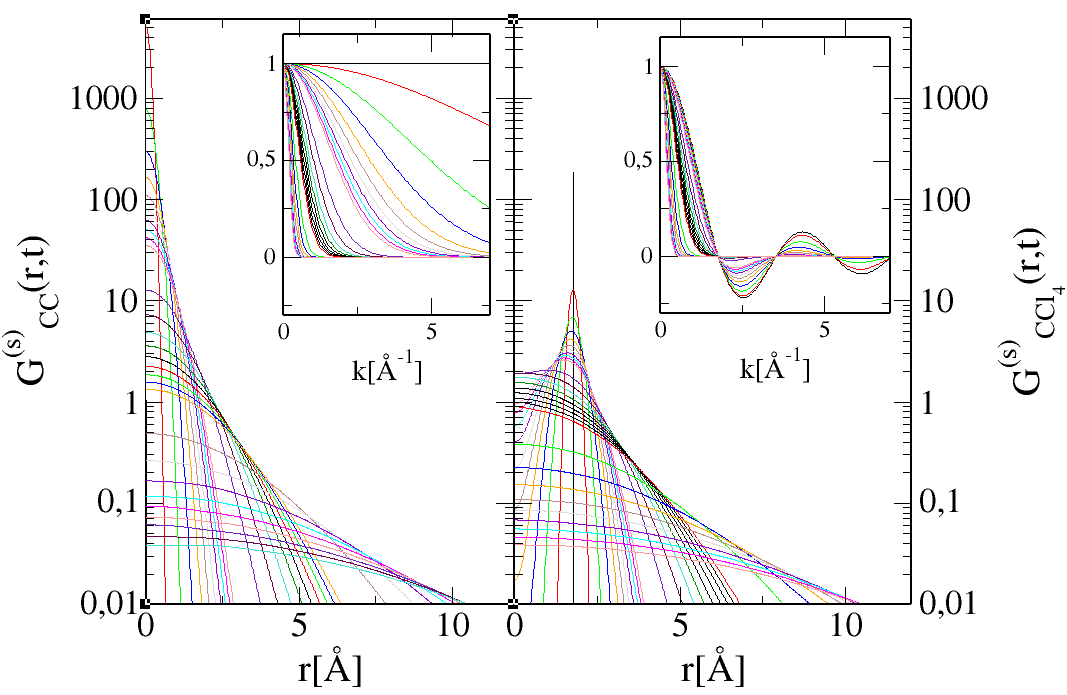}

\caption{CCl\protect\textsubscript{4}: comparison of Self correlations $G^{(s)}(r,t)$
and $F^{(s)}(k,t)$ (inset) for carbon-carbon (left panel) and carbon-chloride
(right panel) pairs of sites. The lines correspond to the 29 times
selected (see text for curve plotting conventions). Note the vertical
log-scale for the main panel.}

\label{Fig-GsFs-CCL4}
\end{figure}

Fig.\ref{Fig-GsFs-CCL4} shows the carbon-carbon self van Hove function
$G_{CC}^{(s)}(r,t)$ in the left panel and the corresponding self-intermediate
scattering function $F_{CC}^{(s)}(k,t)$ in the inset, as well as
the corresponding functions for the carbon tetrachloride dynamical
correlations in the right panel for $G_{CCl_{4}}^{(s)}(r,t)$ and
its inset for $F_{CCl_{4}}^{(s)}(k,t)$. It can be seen that the time
decays are always monotonous in both representations, representing
the spreading of the curves as function of time.

The most notable features are the Dirac delta functions, one at $r=0$
for $G_{CC}^{(s)}(r,t=0)$ (left panel), whose Fourier transform (FT)
is the black horizontal line in the inset, and the other at $r_{d}=2\mathring{A}$
for the C-Cl intermolecular distance (right panel), whose FT is $j_{0}(kr_{d})$
leads to the marked oscillations observed in the inset. Another interesting
feature is that, although the $G^{(s)}(r,t)$ look Gaussian-like,
they cannot be well fitted to Gaussians. This feature invalidates
the usual Gaussian approximation $G^{(s)}(r,t)=\exp(-r^{2}/(4Dt)/(4\pi Dt)^{3/2}$
\cite{Textbook_Hansen_McDonald,Textbook_Boon_Yip,Textbook_Berne_Pecora}.
This is equally true of the small-$k$ Gaussian approximation of $F^{(s)}(k,t)\approx\exp(-Dk^{2}t)$.

\begin{figure}
\includegraphics[scale=0.45]{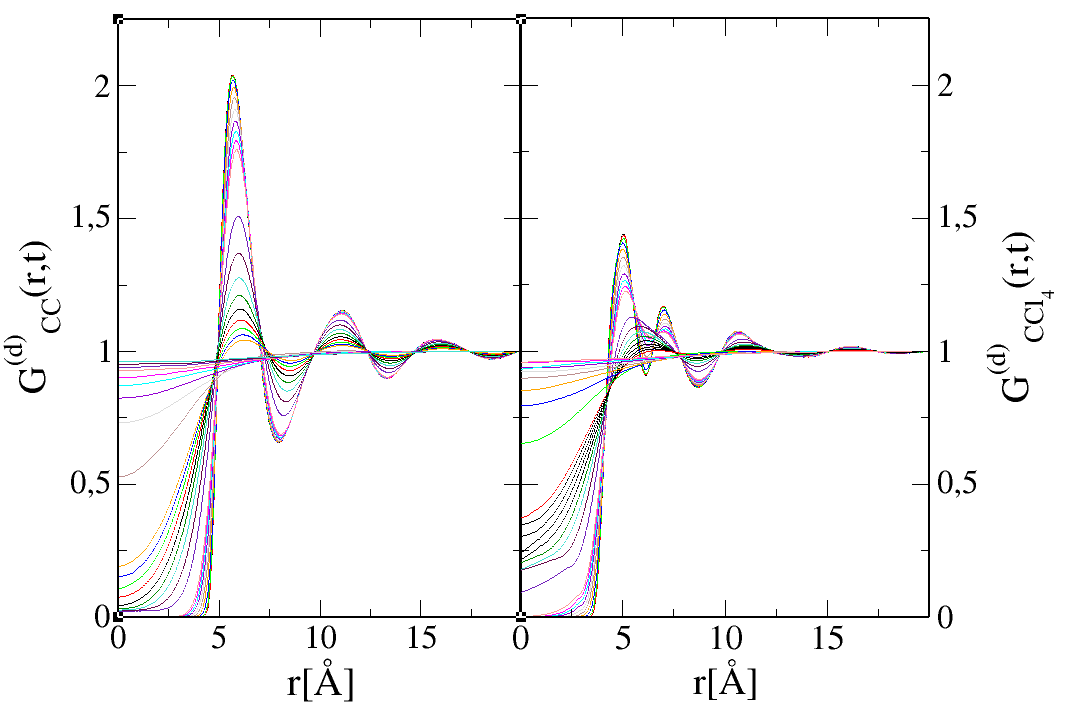}

\caption{CCl\protect\textsubscript{4}: comparison of distinct van Hove functions
$G^{(d)}(r,t)$ for carbon-carbon (left panel) and carbon-chloride
(right panel) pairs of sites. }

\label{Fig-Gd-CCL4}
\end{figure}

Fig.\ref{Fig-Gd-CCL4} shows the distinct part of the van Hove function
$G^{(d)}(r,t)$ between the carbon atoms (left panel) and the carbon-tetrachloride
in the right panel. Basically, these curves look like usual $g(r)$,
but damped in time, as all the curves relax to 1 with time.

The most notable feature is the shift of the first peak of $G_{CCl_{4}}^{(d)}(r,t)$
to larger $r$-values after $t=1$ps, while the second peak is totally
damped. It indicates a remarkable short time decorrelation of the
central carbon with respect to the 4 LJ sites within this time range
of $1$ps. It is equally worth noting that the replacement of the
central particle at the core is not fully relaxed even after 100ps.
It shows a slow diffusion of the molecule from its initial position.

\begin{figure}
\includegraphics[scale=0.45]{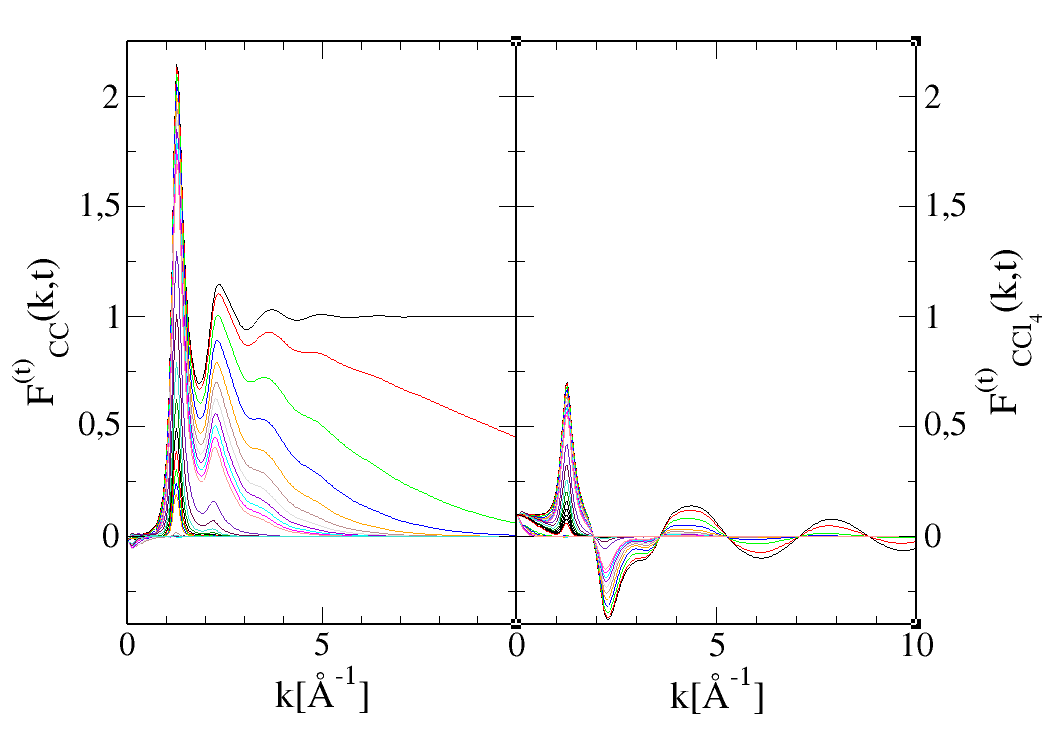}

\caption{CCl\protect\textsubscript{4}: comparison of total intermediate scattering
functions $F^{(s)}(k,t)$ for carbon-carbon (left panel) and carbon-chloride
(right panel) pairs of sites. }

\label{Fig-Ft-CCL4}
\end{figure}

Fig.\ref{Fig-Ft-CCL4} shows the total intermediate scattering function
$F^{(t)}(k,t)$ for the same 2 previous cases. These curves show remarkable
features. First of all, we notice that at $t=0$, $F_{CC}^{(t)}(k,t=0)$
is exactly the static structure factor $S_{CC}(k)$, with the standard
definition $S_{ab}(k)=\delta_{ab}+\rho\int d\mathbf{r}\exp(i\mathbf{k.r)}\left[g_{ab}(r)-1\right]$.
However, in the present case, the asymptote term $\delta_{ab}$ is
related to the intra-molecular correlation, and since $G^{s}(r,t=0)=\delta(r)/\rho$,
it is this term which gives the $\delta_{ab}$ term, which is the
asymptote 1 in case of the CC correlations. 

The most important feature is that the plot shows clearly the decay
of the asymptote with time. The large oscillations in the $F_{CCl_{4}}^{(t)}(k,t)$
are coming from the same self part $j_{0}(kr_{d})$ that is shown
in the right inset of Fig.\ref{Fig-GsFs-CCL4}.

\begin{figure}
\includegraphics[scale=0.45]{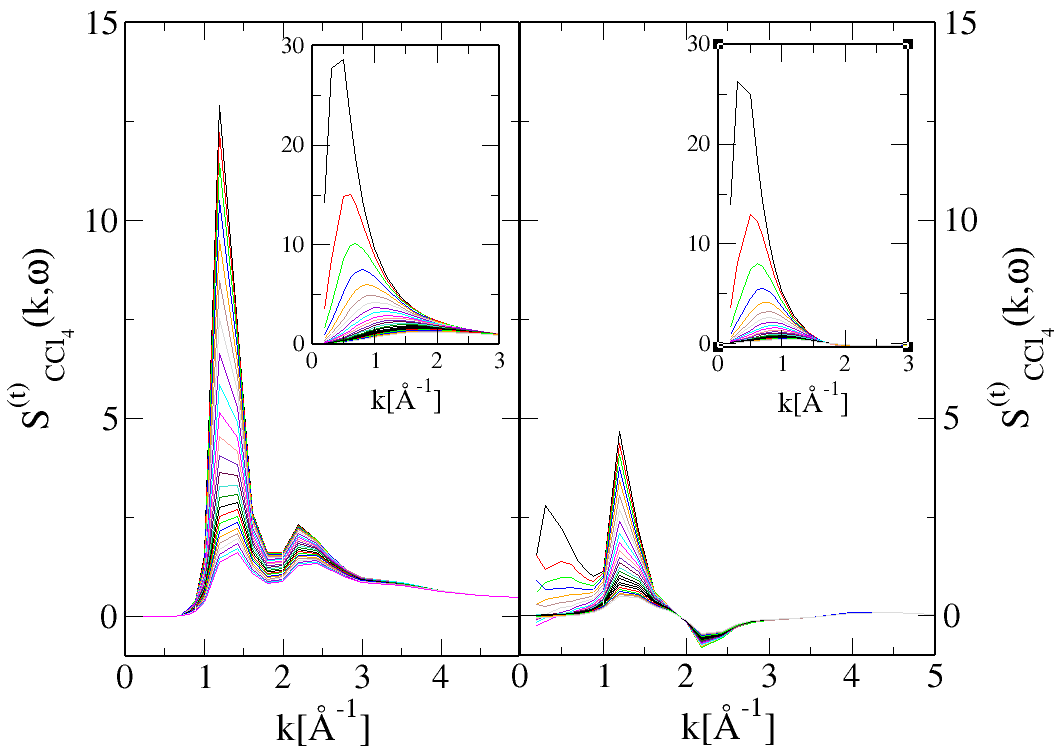}

\caption{CCl\protect\textsubscript{4}: comparison of total dynamical structure
factors $S^{(s)}(k,\omega)$ and corresponding self parts $S^{(s)}(k,\omega)$
(inset) for carbon-carbon (left panel) and carbon-chloride (right
panel) pairs of sites. The lines correspond to the different $\omega=2\pi/t$
frequencies corresponding to the time grid in the $G(r,t)$ plots.}

\label{Fig-Sk-CCL4}
\end{figure}

Fig.\ref{Fig-Sk-CCL4} show the total dynamical structure factor $S^{(t)}(k,\omega)$
and the self part in the inset, for the same atoms as above. The frequencies
$\omega$ correspond to the times on the time grid described earlier,
using the formula $\omega=2\pi/t$, and they are in units of ps$^{-1}$
which is also Ghz. The explicit values are There are 2 remarkable
features. The first is that the main peak of $S(k,\omega)$ coincides
with that of $F(k,t)$ in both cases. The second feature is that the
self part looks very much like the usual Lorentzian approximation
\cite{Textbook_Hansen_McDonald} $S^{(s)}(k,\omega)=Dk^{2}/\pi(\omega^{2}+(Dk^{2})^{2})$,
but cannot be fully fitted to this simple form.

\subsubsection{Polar liquid: acetone}

Acetone is polar molecule, with dipole moment of $10\times10^{20}$C$\mathring{A}$.
Yet, there is no micro-structure formation in this liquid, other than
the usual fluctuations, some related in part to dipole correlations.
We can consider this liquid as a simple disorder liquid.

\begin{figure}
\includegraphics[scale=0.45]{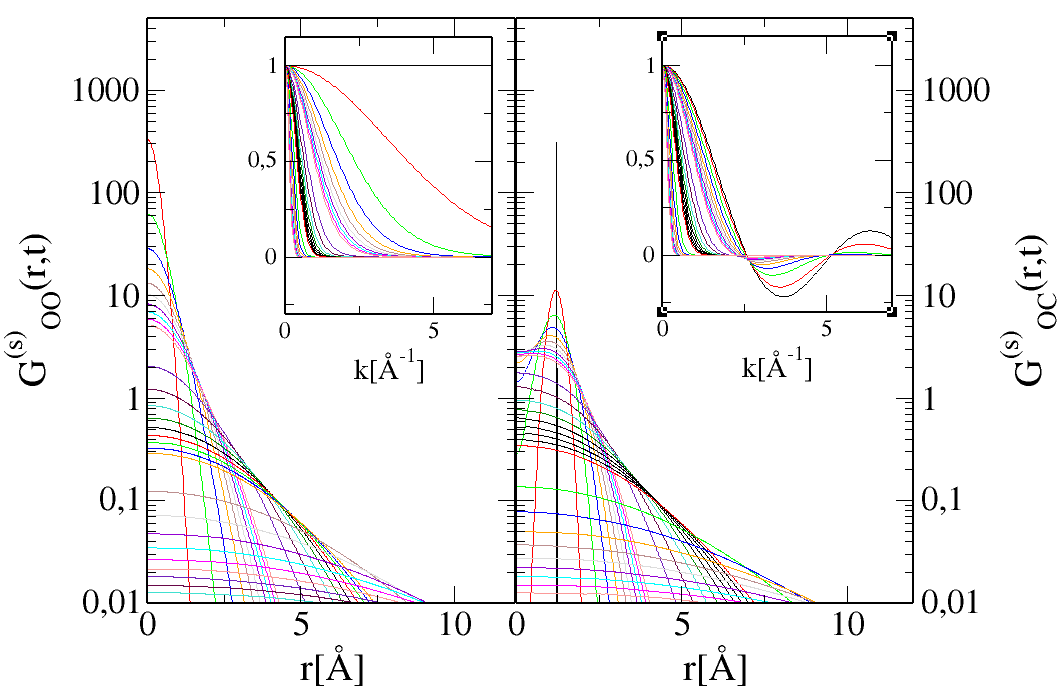}

\caption{Acetone: comparison of Self correlations $G^{(s)}(r,t)$ and $F^{(s)}(k,t)$
(inset) for oxygen-oxygen (left panel) and oxygen-carbon (right panel)
pairs of sites. The lines correspond to the 29 times selected (see
text for curve plotting conventions). Note the vertical log-scale
for the main panel.}

\label{Fig-GsFs-ACE}
\end{figure}

Fig.\ref{Fig-GsFs-ACE} shows the oxygen-oxygen self van Hove function
$G_{OO}^{(s)}(r,t)$ in the left panel and the corresponding self-intermediate
scattering function $F_{OO}^{(s)}(k,t)$ in the inset, as well as
the corresponding functions for the oxygen-carbon dynamical correlations
in the right panel for $G_{OC}^{(s)}(r,t)$ and its inset for $F_{OC}^{(s)}(k,t)$.
These curves have strong similarities with those for CCL4 in Fig.\ref{Fig-GsFs-CCL4}. 

They also show characteristic differences. The $r$-decay and the
time decay in $r$-space are faster for acetone than for CCl\textsubscript{4}.
However, since all $G^{(s)}(r,t)$ integrate to 1 (that is $F^{(s)}(k=0,t)=1),$it
means that self correlations are longer ranged for acetone than CCl\textsubscript{4},
and especially at larger times. We can attribute this to dipole correlation
with neighbours. However, it can be seen from the inset that $F_{OO}^{(s)}(k,t)$
decay faster for acetone than for CCl\textsubscript{4}. We speculatively
attribute this to CCl\textsubscript{4} being more ``spherical''
than acetone, possibly because of the dipole moment of the latter.

\begin{figure}
\includegraphics[scale=0.45]{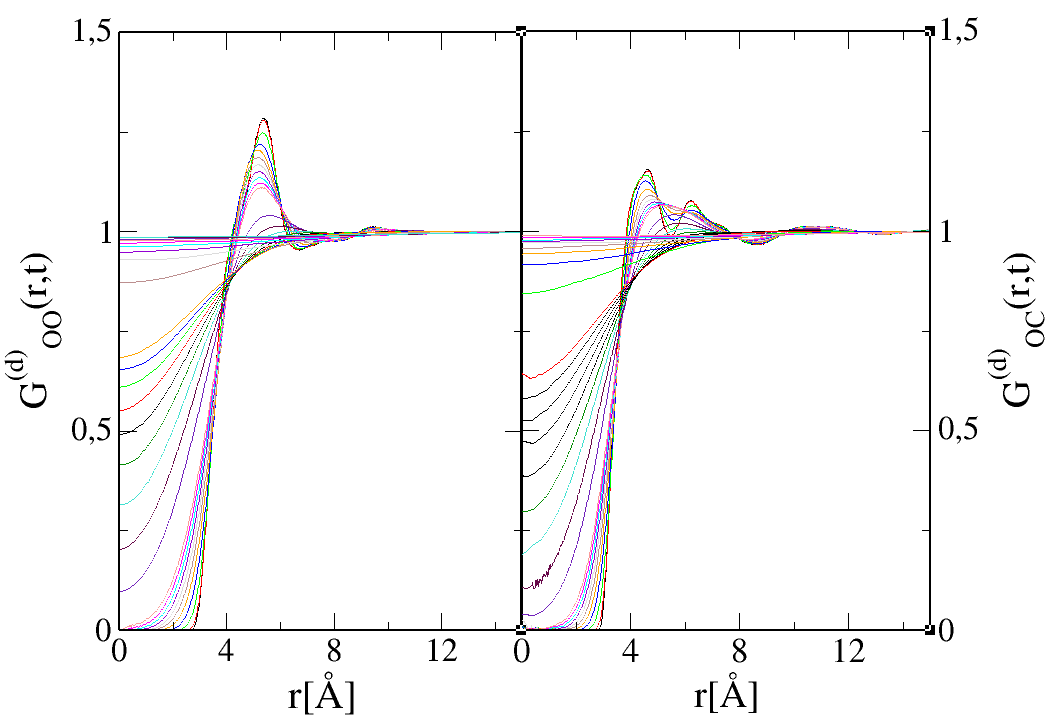}

\caption{Acetone: comparison of distinct van Hove functions $G^{(d)}(r,t)$
for oxygen-oxygen (left panel) and oxygen-carbon (right panel) pairs
of sites.}

\label{Fig-Gd-ACE}
\end{figure}

Fig.\ref{Fig-Gd-ACE} shows the time decay of distinct van Hove correlations
for OO and OC. These are basically the equivalent of time dependent
pair correlation functions since $g(r)=G^{(d)}(r,t=0)$, which is
why they are worth analyzing. The comparison with CCl\textsubscript{4}
in Fig.\ref{Fig-Gd-CCL4} shows remarkable differences. For acetone,
the main peaks are much smaller and the spatial oscillations much
less pronounced (faster decay of spatial correlations). This is surprising
since both liquids are dense. Again, this can be attributed to dipole
correlations, using the following argument. The maintenance of dipole
alignments can be achieved by several molecular positions, hence,
for entropic reasons, there would be a faster decorrelation between
sites, while the dipole correlations are maintained for longer times.

We note that the first peak spreading with time in the cross site
correlation is very similar to that of CCl\textsubscript{4}.

\begin{figure}
\includegraphics[scale=0.45]{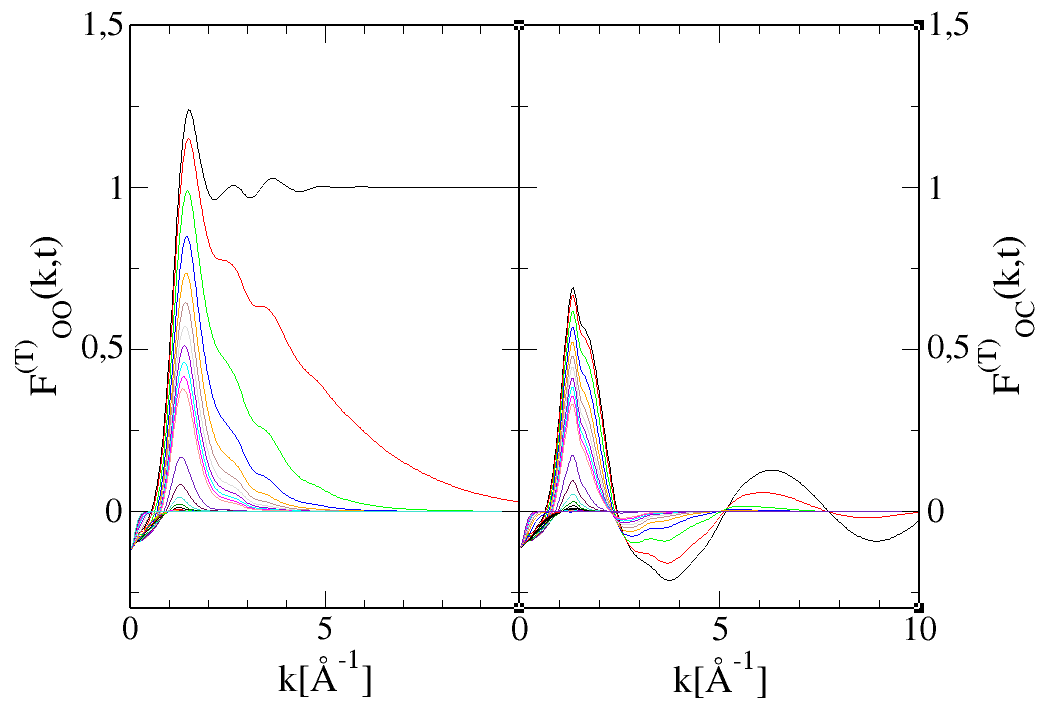}

\caption{Acetone: comparison of total intermediate scattering functions $F^{(s)}(k,t)$
for oxygen-oxygen (left panel) and oxygen-carbon (right panel) pairs
of sites.}

\label{Fig-Fkt-ACE}
\end{figure}

Fig.\ref{Fig-Fkt-ACE} shows the total intermediate scattering function
for the same two pairs of sites. We observe the same faster temporal
decay for acetone thane for CCl\textsubscript{4}, as that observed
for the self part in the insets of Fig.\ref{Fig-GsFs-ACE}. 

The oscillations of the OC cross-correlations are also larger for
acetone than for CCl\textsubscript{4}. But this can be interpreted
as a Fourier transform mathematical consequence of the core and first
peak parts of the $G^{(d)}$function: the higher the first peak, the
tighter the oscillations in $k$-space.

\begin{figure}
\includegraphics[scale=0.45]{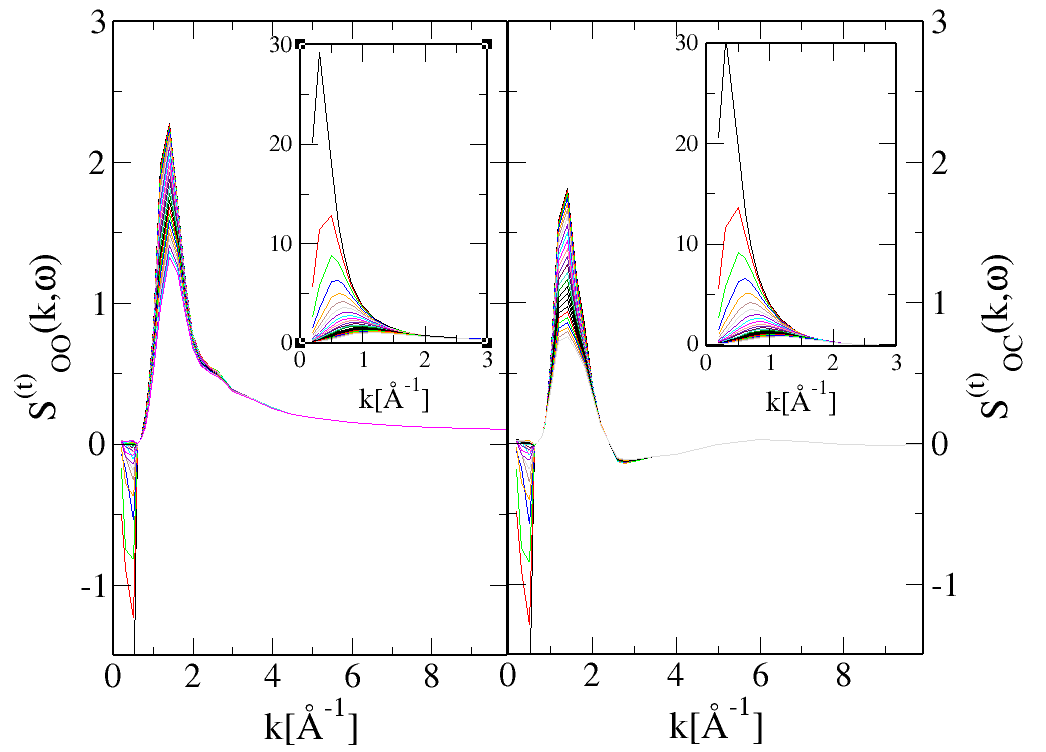}

\caption{Acetone: comparison of the total dynamical structure factors $S^{(s)}(k,\omega)$
and corresponding self parts $S^{(s)}(k,\omega)$ (inset) for oxygen-oxygen
(left panel) and oxygen-carbon (right panel) pairs of sites. }

\label{Fig-Sk-ts-ACE}
\end{figure}

Fig.\ref{Fig-Sk-ts-ACE} shows the total and self(inset) dynamical
structure factors. The most prominent differences with CCl\textsubscript{4}
of Fig.\ref{Fig-Sk-CCL4} are the absence of second peak around $k\approx2\mathring{A}$,
and the small negative part near $k\approx0$. The negative part is
in fact a numerical artifact coming from the small negative part near
$k=0$ of the $F(k,t)$ in Fig.\ref{Fig-Fkt-ACE}. This negative part
is coming from a known problem in g(r), hence G(r,t), asymptote in
simulations: they do not tend exactly to 1, but slightly below \cite{Lebowitz1961}.
This problem is well known for the Kirkwood-Buff calculation in computer
simulations \cite{Perera2004,Ganguly2013,Krueger2013,Milzetti2018,Dawass2018}.

Aside from these, both sets of curves for CCl\textsubscript{4} and
acetone look quite similar.

\subsection{Complex disorder liquids}

Herein, we study water and ethanol, which are both archetypical examples
of hydrogen bonded liquids. We do not expect the self parts to show
much differences, unless indirectly, since most of differences are
related to the distinct part which contain the specific inter-molecular
parts which make the difference between strongly and weakly clustered
liquids.

\subsubsection{Water}

\begin{figure}
\includegraphics[scale=0.45]{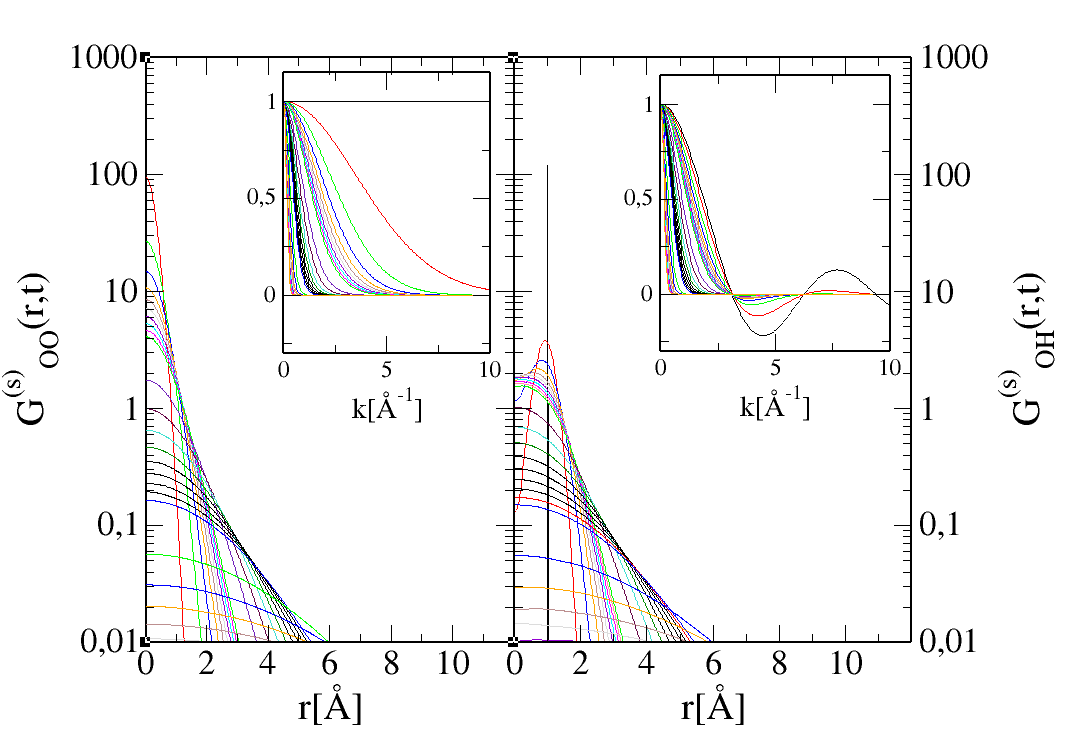}

\caption{Water: comparison of Self correlations $G^{(s)}(r,t)$ and $F^{(s)}(k,t)$
(inset) for oxygen-oxygen (left panel) and oxygen-hydrogen (right
panel) pairs of sites. The lines correspond to the 29 times selected
(see text for curve plotting conventions). Note the vertical log-scale
for the main panel.}

\label{Fig-GsFs-WAT}
\end{figure}

Fig.\ref{Fig-GsFs-WAT} shows the self van Hove correlations both
in $r$ and $k$(inset) space. The comparison with the simple disorder
counter parts of CCl\textsubscript{4} and acetone in Figs{[}\ref{Fig-GsFs-CCL4},\ref{Fig-GsFs-ACE}{]},
shows that there is quite a bit of resemblance with both of them.
Perhaps the most importance difference is the smaller height of the
main peaks, both at $r=0$ and $r=1\mathring{A}$ , and the faster
decay in $r$-space. This is an important feature, since, because
$F^{(s)}(k=0,t)=1$, it implies the existence of longer ranged $r$-correlations
for water, which is an indirect proof of the Hbond network persistence
for this particular liquid, over their absence for simple disorder
liquids. Despite this difference, we observe that the $F^{(s)}(r,t)$
are quite similar. 

This is, speculatively, an indication that water is a 2-state liquid:
it appears as a r-space long range correlated liquid, and at the same
time a disordered one in k-space. This is probably an typical difference,
specially when comparing with the $F^{(s)}$ for acetone in Fig.\ref{Fig-GsFs-ACE},
which also appear as long ranged in r-space (as discussed previously
sub-section 4.1.2), but differs in k-space.

\begin{figure}
\includegraphics[scale=0.45]{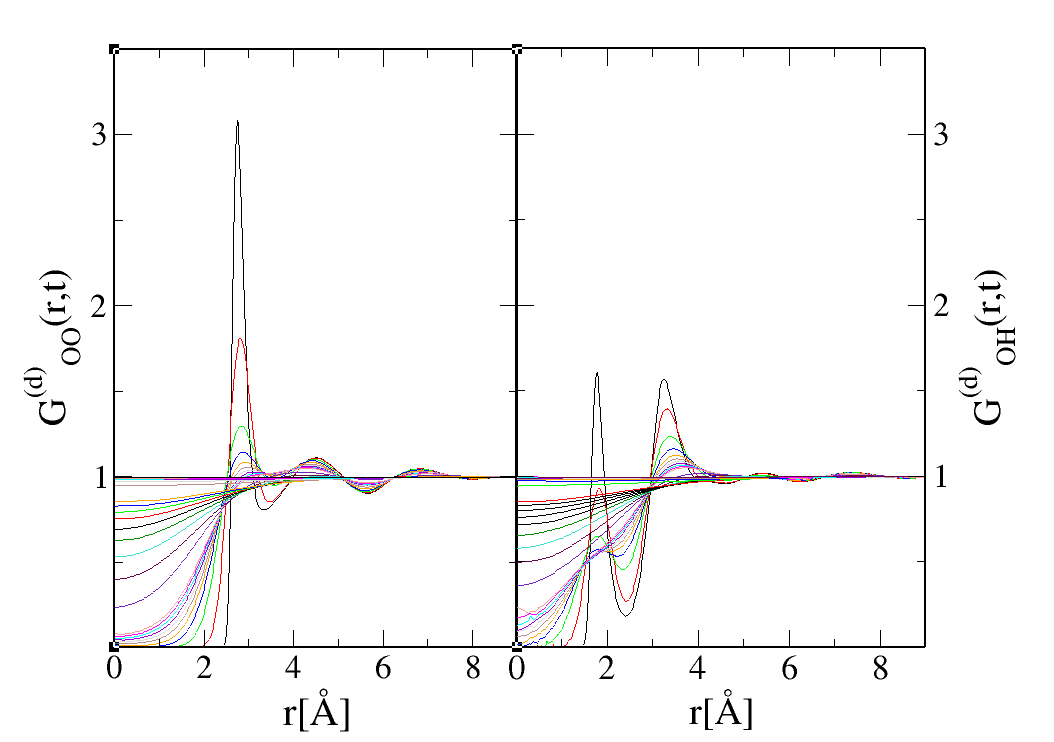}

\caption{Comparison of distinct van Hove functions $G^{(d)}(r,t)$ for oxygen-oxygen
(left panel) and oxygen-hydrogen (right panel) pairs of sites.}

\label{Fig-Gd-WAT}
\end{figure}

Fig.\ref{Fig-Gd-WAT} shows the distinct dynamical correlations. Since
$g_{OO}(r)=G_{OO}^{(d)}(r,t=0)$ we can observe the typical pair correlation
of water, which differs considerably from that of simple disorder
liquids, such as illustrated by CCl\textsubscript{4} and acetone,
and its decay with time. The most striking difference is the rapid
decay of the O-O contact peak at $r\approx3\mathring{A}$ and the
O-H Hbond peak at $r\approx2\mathring{A}$ . This appears at first
quite contradictory, since on would think that the Hbond related peaks
should show more persistence than the contact peak of simple disorder
liquids. This in fact similar to that observed for the self-functions
above: what is important is not how rapidly the Hbond correlations
are lost, but how rapidly they are reestablished. The latter is better
observed in k-space (see below). Nevertheless, it appears that Hbonding
has a very short life time since it decays very fast, which we have
already observed in our previous work \cite{Jukic2021}.

\begin{figure}
\includegraphics[scale=0.45]{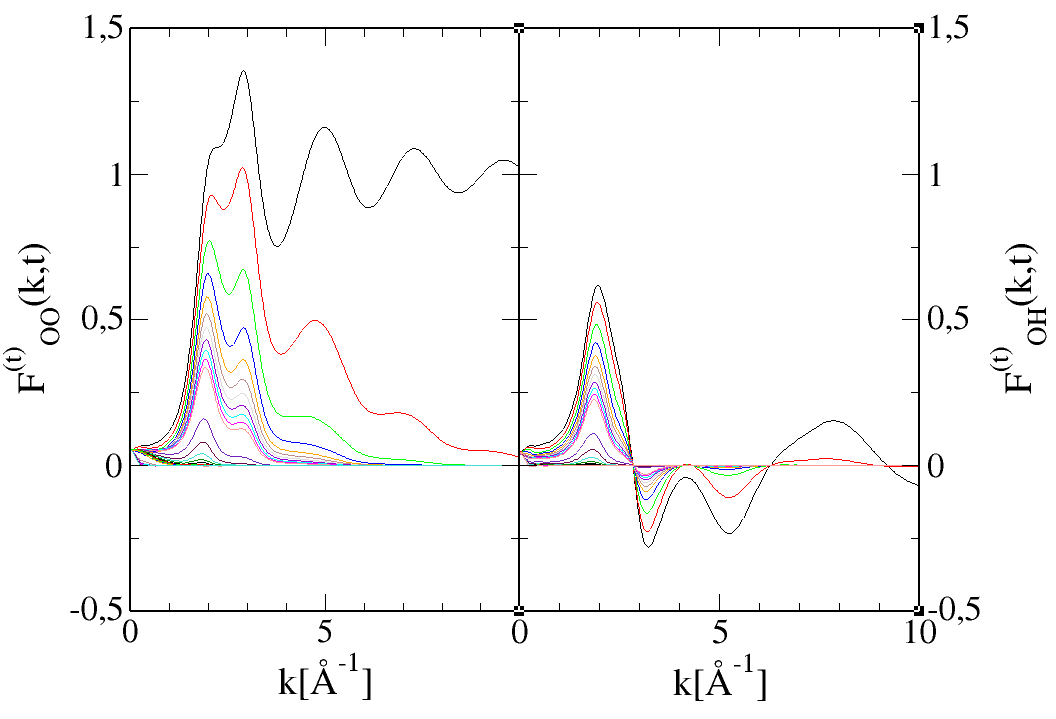}

\caption{Water: comparison of total intermediate scattering functions $F^{(s)}(k,t)$
for oxygen-oxygen (left panel) and oxygen-hydrogen (right panel) pairs
of sites.}

\label{Fig-Ft-WAT}
\end{figure}

The above analysis is further confirmed in Fig.\ref{Fig-Ft-WAT},
which shows the total intermediate scattering functions. We observe
the typical split-peak feature of the structure factor of water, since
$S_{OO}(k)=F_{OO}^{(t)}(k,t=0)$. But what seems interesting is that
the shoulder peak at $k\approx2\mathring{A}^{-1}$ ($r\approx3\mathring{A}$)
decays more slowly than the Hbond peak at $k\approx3\mathring{A}^{-1}$($r\approx2\mathring{A}$).
We have recently proposed \cite{Lovrincevic_ChemRXiv} that the shoulder
peak is in fact the cluster peak for water, which points to water
as forming mostly a dimer, which is a direct consequence of charge
order. This is further confirmed by the fact that at $r\approx3\mathring{A}$we
observe an anti-peak in the OH correlations.

Interestingly, the asymmetry of the time decay of these 2 peaks is
very strongly reminiscent of their temperature dependence \cite{WaterTemperature}:
the Hbond peak is lost more quickly by heating than the contact-cluster
peak.

\begin{figure}
\includegraphics[scale=0.45]{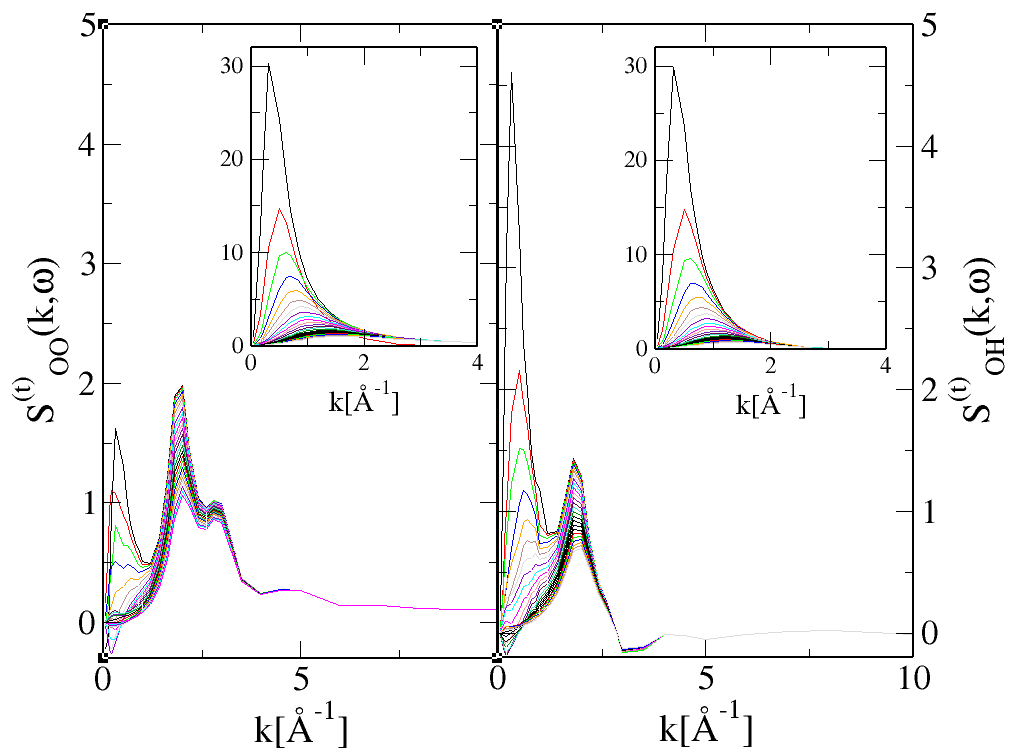}

\caption{Water: comparison of the total dynamical structure factors $S^{(s)}(k,\omega)$
and corresponding self parts $S^{(s)}(k,\omega)$ (inset) for oxygen-oxygen
(left panel) and oxygen-hydrogen (right panel) pairs of sites.}

\label{Fig-Sk-ts-WAT}
\end{figure}

Fig.\ref{Fig-Sk-ts-WAT} shows the dynamical structure factors for
the OO and OH atom pairs. The most striking feature is the small-k
peak at $k\approx0.3\mathring{A}^{-1}$, which corresponds to $r\approx20\mathring{A}$.
This peak is even more prominent for the OH correlations, indicating
that its origin is mostly from the Hbond correlations. This peak is
more important than dual peak that we observe at $k\approx2\mathring{A}^{-1}$
and $k\approx3\mathring{A}^{-1}$, which are directly related to those
in Fig.\ref{Fig-Ft-WAT}. In fact, this new peak is the dynamical
equivalent of static pre-peak in the static structure factor.

The distance size of $20\mathring{A}$ corresponds to a radius of
$10\mathring{A}$, which is about the range of the marked oscillations
in the $g_{OO}(r)$: the ``3 peak structure'' discussed in our previous
work \cite{Perera2011}. It this represent a kinetic clustering, as
opposed to the static clustering which is observed through the pre-peak
of the static structure factor. This analysis is further confirmed
in the the case of ethanol below.

\subsubsection{Ethanol}

Contrary to water, alcohols have a well defined scattering pre-peak
\cite{EXP_Narten_MethEth,EXP_Sarkar_Methanol,EXP_Sarkar_ethanol,EXP_Finns_alcohols,EXP_Alcohols_Karmakar1999,EXP_Matija_alcohols,EXP_Pusztai_MethEthProp,EXP_Prop_Sillren2013,EXP_SIM_Matija_Butanol}.
Herein, we focus on the case of ethanol for illustrative purposes. 

\begin{figure}
\includegraphics[scale=0.45]{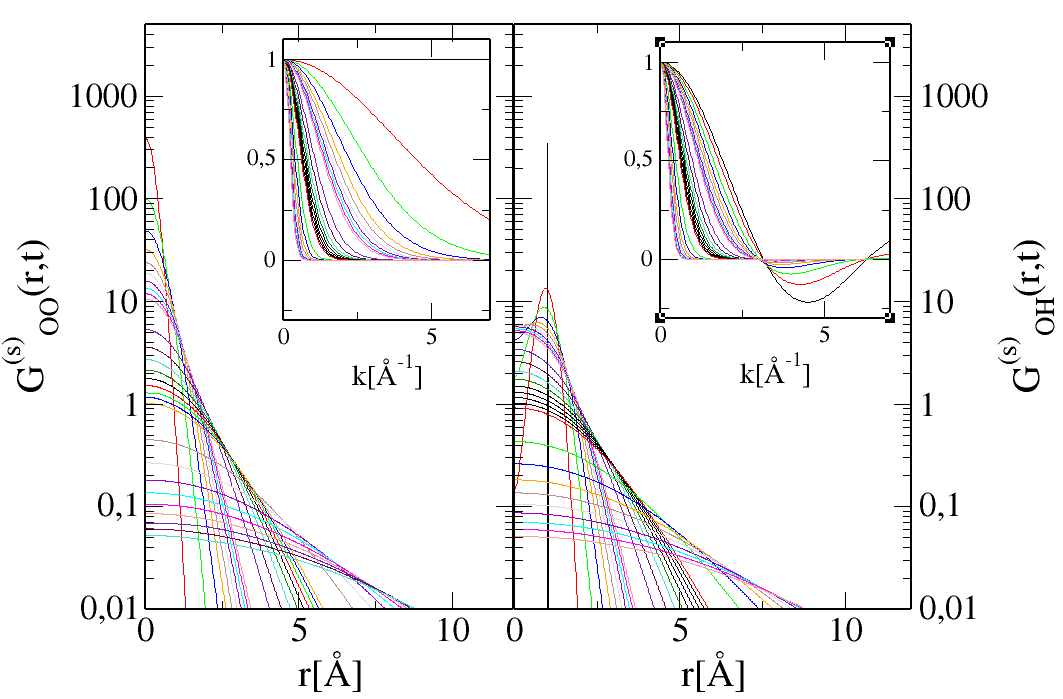}

\caption{Ethanol: comparison of Self correlations $G^{(s)}(r,t)$ and $F^{(s)}(k,t)$
(inset) for oxygen-oxygen (left panel) and oxygen-hydrogen (right
panel) pairs of sites. The lines correspond to the 29 times selected
(see text for curve plotting conventions). Note the vertical log-scale
for the main panel.}

\label{Fig-GsFs-ETH}
\end{figure}

Fig.\ref{Fig-GsFs-ETH} shows the self parts of van Hove and intermediate
scattering (insets) functions for OO and OH correlations. The general
features are not so much different than that observed form the simple
disorder liquids, which gives the false impression that the decorrelation
of a single ethanol molecule is similar to that of CCl\textsubscript{4}
or acetone.

This is quite interesting, in the sense that water stands apart, even
though both water and alcohols are Hbonded liquids, since we already
observed distinct features at the level of the self dynamical correlations
for this particular liquid.

\begin{figure}
\includegraphics[scale=0.45]{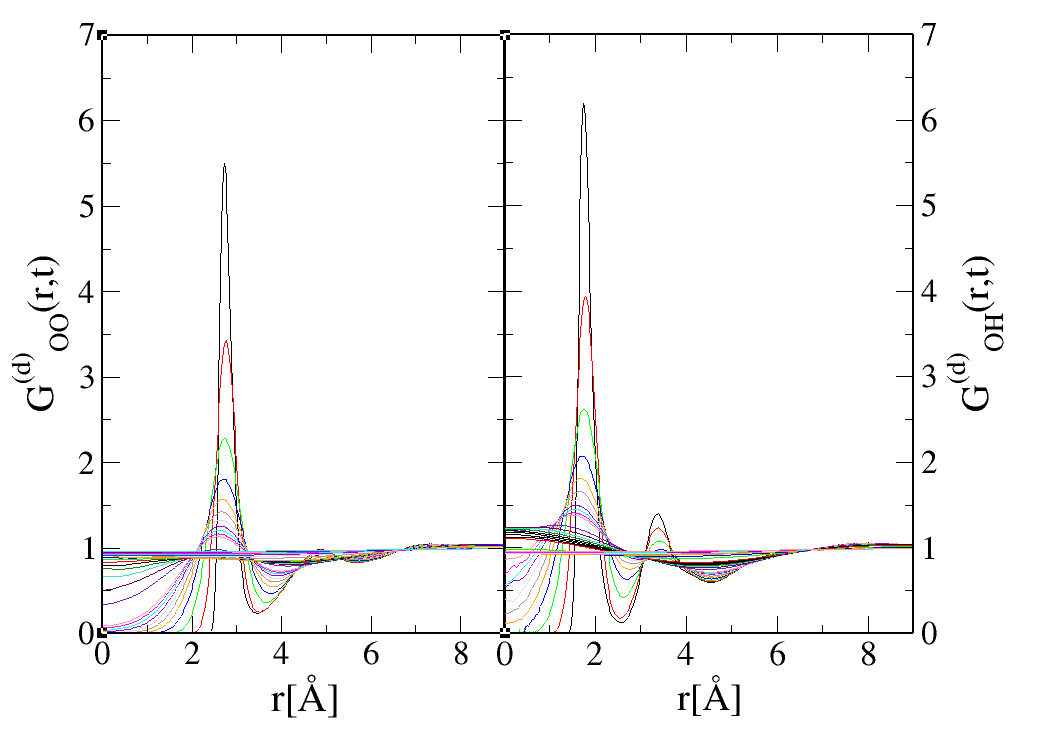}

\caption{Ethanol: comparison of distinct van Hove functions $G^{(d)}(r,t)$
for oxygen-oxygen (left panel) and oxygen-hydrogen (right panel) pairs
of sites.}

\label{Fig-Gd-ETH}
\end{figure}

In order to find the specificity related to Hbonding, we turn towards
the distinct correlation functions in Fig.\ref{Fig-Gd-ETH}. We observe
several similarities with water. The high first peak is a witness
of Hbonding, just like water, and it is also seen to decay very fast.
A specificity of alcohols is the depletion correlations after the
main peak, which are at the origin of the alcohol pre-peak \cite{Pozar2016,Pozar2020}.

These depletion correlation of the second and higher neighbours are
related to the chain formation of the hydroxyl groups, and express
the fact that the corresponding neighbours are in reduced number with
respect to full space filling. It is seen that the depleted region
re-emerge above 1 around $r\approx7-8\mathring{A}.$ These depletion
correlations are seen to decay more slowly, a feature more visible
in the Fourier transform discussed below.

\begin{figure}
\includegraphics[scale=0.45]{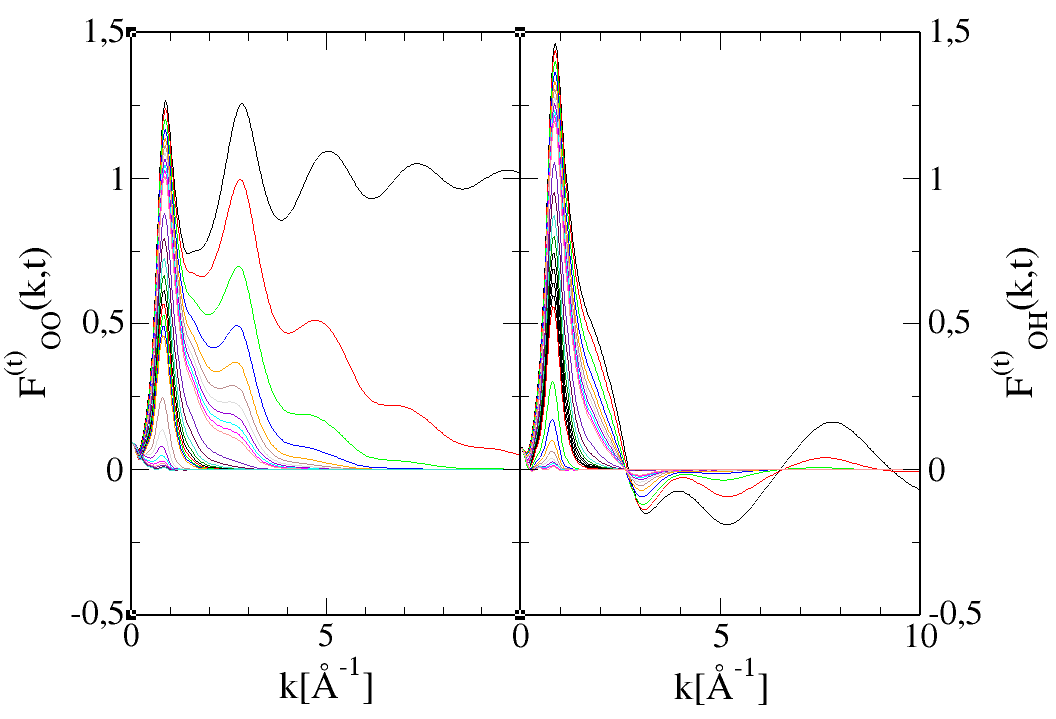}

\caption{Ethanol: comparison of total intermediate scattering functions $F^{(s)}(k,t)$
for carbon-carbon (left panel) and carbon-chloride (right panel) pairs
of sites.}

\label{Fig-Ft-ETH}
\end{figure}

Fig.\ref{Fig-Ft-ETH} shows how the pre-peak at $k_{PP}\approx0.75-0.8\mathring{A}^{-1}$
($r\approx7-8\mathring{A}$) and Hbond peak at $k_{MP}\approx3\mathring{A}^{-1}$
($r\approx2\mathring{A}$) of ethanol decay in time. $k_{PP}$ corresponds
to the depletion range observed in Fig.\ref{Fig-Gd-ETH}, while the
main peak is exactly that of water and corresponds to the O-H Hbonding
distance. Just like water, the cluster pre-peak is seen to decay much
more slowly than the main peak. The rate of decay is seen to be even
slower than the dimer cluster peak of water. 

This is in line with the fact that alcohol have long lived chain clusters
\cite{Fujii2018,Jukic2021}.

\begin{figure}
\includegraphics[scale=0.45]{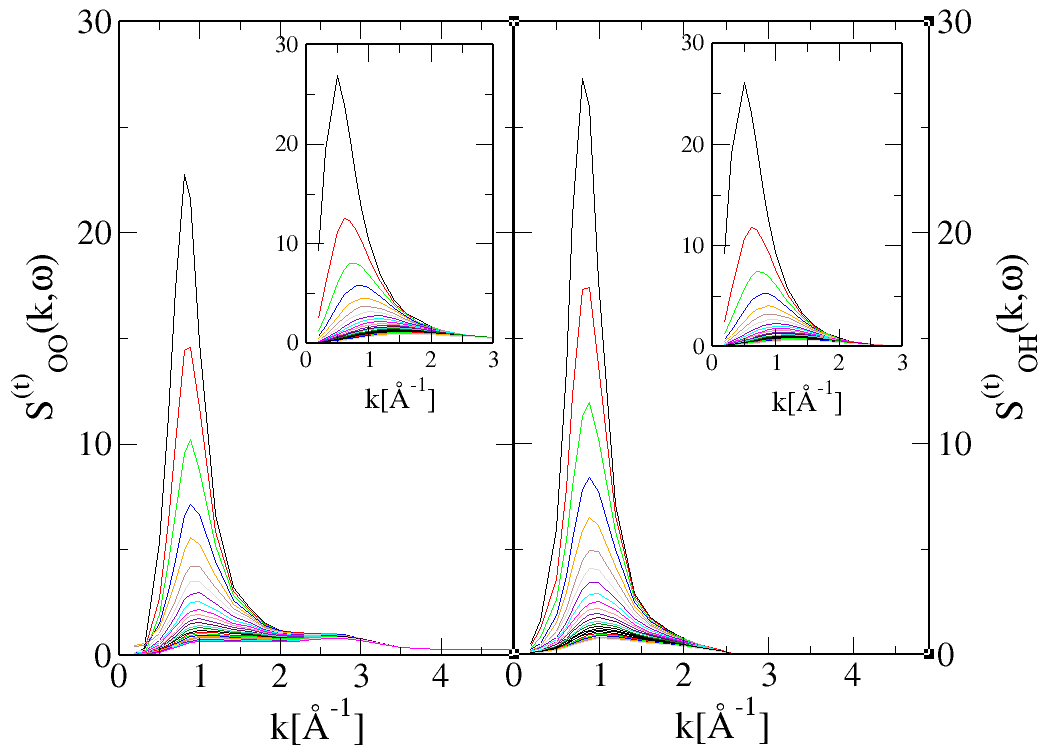}

\caption{Ethanol: comparison of the total dynamical structure factors $S^{(s)}(k,\omega)$
and corresponding self parts $S^{(s)}(k,\omega)$ (inset) for oxygen-oxygen
(left panel) and oxygen-hydrogen (right panel) pairs of sites.}

\label{Fig-Sk-ts-ETH}
\end{figure}

Fig.\ref{Fig-Sk-ts-ETH} shows the dynamical structure factors for
the OO and OH atom pairs. We observe a very high dynamical pre-peak
at $k\approx0.7-0.8\mathring{A}^{-1}$, similar to water, but much
higher in magnitude, which corresponds exactly to the static pre-peak
of ethanol observed above. In contrast to the case above, we see that
the peak at $k\approx3\mathring{A}$ is much smaller. These features
highlight the dynamics of cluster formation in ethanol, and more generally
in alcohols. 

This analysis reveals that the dynamical structure factors are better
indicators of the cluster dynamics than the static structure factors,
and highlight the underlying kinetics.

\section{Conclusion}

In this work, we have conducted an analysis of the dynamics of liquids
are revealed by the pair correlation functions, both in $r$ and $k$
space, as well in time and frequency. To this end, the van Hove, intermediate
scattering functions and dynamical structure factors have been calculated
in computer simulations of model liquids. Having in mind the structural
differences between simple disorder liquids and clustering liquids,
we have studied CCl\textsubscript{4} and acetone for the first case,
water and ethanol for the second case. Our study reveals that many
important structural differences between the two categories of liquids
exist, as exemplified by the analysis of the various dynamical pair
correlation functions. The most important difference is the existence
of kinetic processes due to clustering, which are prominently found
in the small $k$ part of atom-atom dynamical structure factors $S(k,\omega)$
of complex disorder liquids, and are totally absent from the simple
disorder liquids. This feature is not so much perceptible from the
static structure factor $S(k)$ and intermediate scattering functions
$F(k,t)$, as exemplified for the case of water. Since the x-ray radiation
scattering experiments are related to the static structure factors,
it appear important to refer to neutron scattering experiments in
order to gain access to the dynamical scattering intensity. We expect
that this work will motivate research in this direction.

\section*{Acknowledgments}

This work has been supported in part by the Croatian Science Foundation
under the project UIP-2017-05-1863 ``Dynamics in micro-segregated
systems''.

\bibliographystyle{jpc_title}
\bibliography{dyna2}

\end{document}